\pdfoutput=1
\documentclass[aps,prl,twocolumn,a4paper]{revtex4}
\usepackage[latin1]{inputenc}
\usepackage{graphicx}
\usepackage{amsmath,amssymb}
\usepackage{hyperref}
\hypersetup{colorlinks=true,
						linkcolor=black,
						anchorcolor=black,
						citecolor=black,
						filecolor=black,
						menucolor=black,
						pagecolor=blue,
						urlcolor=black
}
\usepackage{datetime}
\usepackage{upgreek}
\usepackage{placeins}
\usepackage{braket}
\usepackage{units}
\usepackage{xcolor}

\newcommand\cop[2] { \hat c_{{#1}{#2}} }
\newcommand\cdag[2] { \hat c^\dagger_{{#1}{#2}} }

\makeatletter
\newcommand*{\newbibstartnumber}[1]{%
  \apptocmd{\thebibliography}{%
    \global\c@NAT@ctr #1\relax
    \addtocounter{NAT@ctr}{-1}%
  }{}{}%
}
\makeatother

\begin{document}
\title{Supersolid formation in a quantum gas breaking continuous translational symmetry}

\author{Julian L\'eonard}
\author{Andrea Morales}
\author{Philip Zupancic}
\author{Tilman Esslinger}
\email{esslinger@phys.ethz.ch}
\author{Tobias Donner}
\affiliation{Institute for Quantum Electronics, ETH Zurich, 8093 Zurich, Switzerland}

\maketitle

{\bf \noindent 
The concept of a supersolid state is paradoxical. It combines the crystallization of a many-body system with dissipationless flow of the atoms it is built of. This quantum phase requires the breaking of two continuous symmetries, the phase invariance of a superfluid and the continuous translational invariance to form the crystal \cite{Boninsegni2012, Chan2013}. Proposed for helium almost 50 years ago \cite{Andreev1969, Thouless1969}, experimental verification of supersolidity remains elusive \cite{Kim2004, Kim2012}. A variant with only discrete translational symmetry breaking on a preimposed lattice structure, the `lattice supersolid' \cite{Matsuda1970}, has been realized based on self-organization of a Bose-Einstein condensate (BEC) \cite{Baumann2010, Mottl2012}. However, lattice supersolids do not feature the continuous ground state degeneracy that characterizes the supersolid state as originally proposed. Here we report the realization of a supersolid with continuous translational symmetry breaking. The continuous symmetry emerges from two discrete spatial ones by symmetrically coupling a BEC to the modes of two optical cavities. We establish the phase coherence of the supersolid and find a high ground-state degeneracy by measuring the crystal position over many realizations through the light fields leaking from the cavities. These light fields are also used to monitor the position fluctuations in real-time. Our concept provides a route to creating and studying glassy many-body systems with contrallably lifted ground-state degeneracies, such as supersolids in the presence of disorder.
}
\begin{figure}[b]
    \includegraphics[width=\columnwidth]{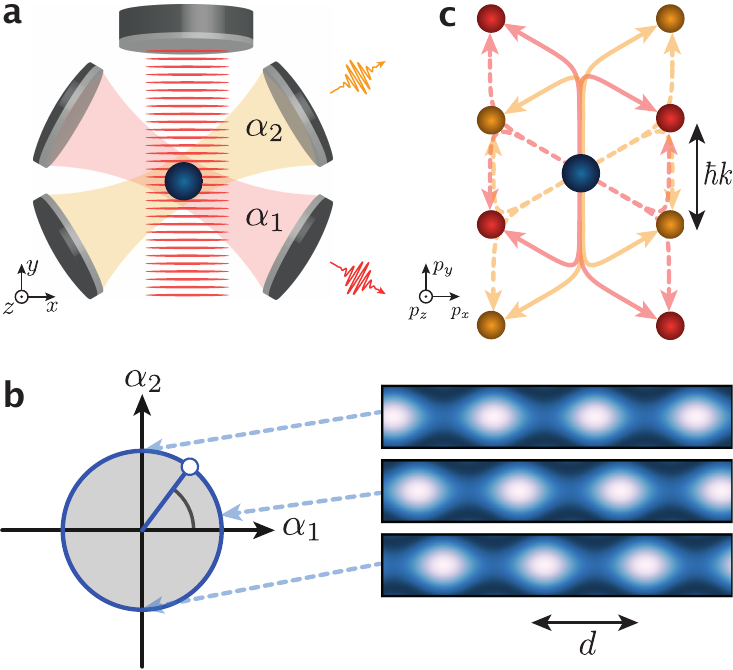}
    \caption{ \textbf{Breaking continuous translational symmetry in a superfluid quantum gas. a,} A BEC (blue) is trapped at the intersection of two optical cavities crossing at an angle of $60^\circ$ and exposed to a one-dimensional optical lattice formed by a transverse pump beam (red stripes). Photons scattered by the atoms populate the cavity mode 1 (2), shaded in red (yellow),  with coherent field amplitude $\alpha_1$ ($\alpha_2$) that can be detected when leaking from the cavity. \textbf{b,}  The two BEC-cavity systems individually exhibit parity symmetry and are combined to a U(1) symmetry. The resulting ground state manifold in terms of the order parameters $\alpha_1$ and $\alpha_2$ forms a circle. The interference potential of the cavity fields with the transverse pump field is $d$-periodic and moves continuously along the $x$--axis when changing the angle in the $\alpha_1$--$\alpha_2$ plane. \textbf{c,} The atomic momentum states associated with coherent scattering processes between transverse pump and cavity 1 (red) and 2 (yellow), starting from the ground state with zero momentum (blue). Solid (dashed) lines describe excitation processes involving the creation (annihilation) of a cavity photon with momentum $\hbar k$.}
    \label{fig1}
\end{figure}

At a phase transition, the type of symmetry that is broken has fundamental consequences on the system. While a discrete symmetry results in disconnected robust states with gapped excitations, a continuous symmetry leads to an infinite number of degenerate ground states that can evolve from one to another without energy cost, making the system highly susceptible to fluctuations. In quantum gas experiments a number of different phase transitions were explored \cite{Bloch2008}, however, engineering a continuous spatial symmetry breaking in a superfluid has remained a challenge. Efforts are being made on long-range interacting dipolar \cite{Lahaye2009, Kadau2016, Zeiher2016, Moses2015} as well as spin-orbit coupled quantum gases \cite{Goldman2014, Li2016}, and light-matter interaction with a mode continuum \cite{Ostermann2016, Gopalakrishnan2009}, where either superfluidity \cite{Labeyrie2014} or the number of modes \cite{Kollar2016} has posed a limitation so far. 

\begin{figure*}[t]
    \includegraphics[width=0.85\textwidth]{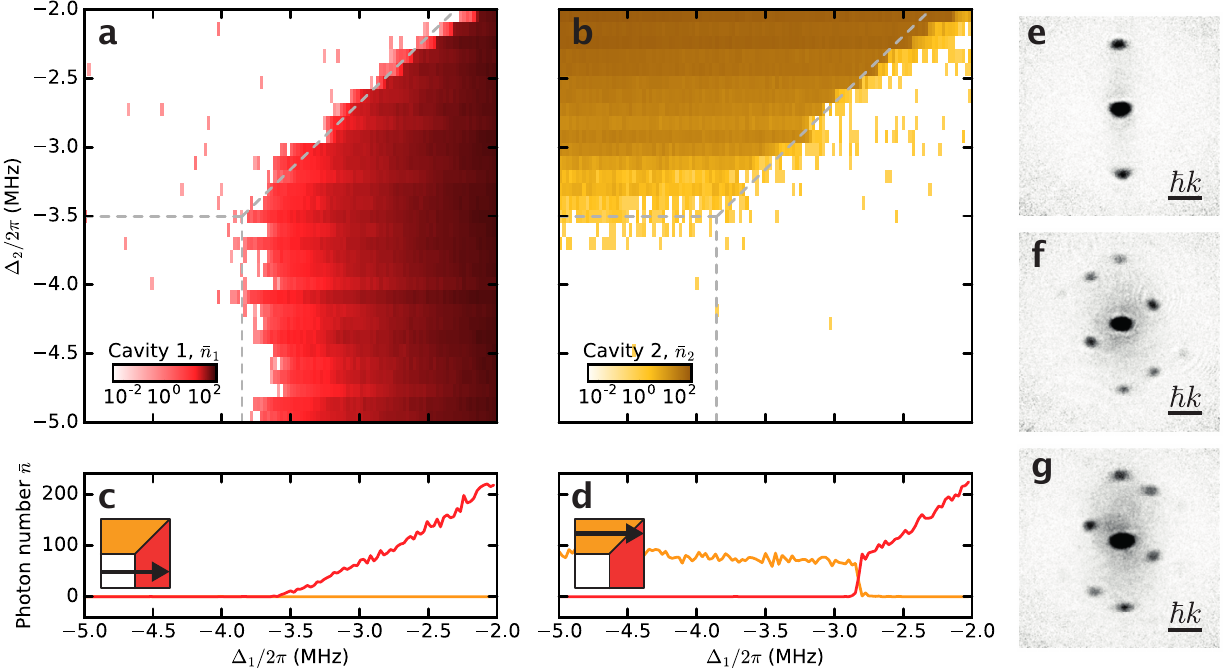}
    \caption{{\bf Self-organized phases in the two cavity--system.} \textbf{a--b,} Mean intracavity photon numbers $\bar{n}_i = | \alpha |^2_i$ as a function of the cavity-pump detunings $\Delta_1$ and $\Delta_2$ for aconstant transverse pump lattice depth of $38(1)\,\hbar\omega_\mathrm{rec}$. Each horizontal line was taken for both cavities simultaneously in a single run of $25\,\mathrm{ms}$. Dashed lines show the phase boundaries expected from a mean-field model (see Methods). \textbf{c--d,} Photon traces for cavity 1 (red) and 2 (yellow) as a function of $\Delta_1$, for $\Delta_2/2\pi=-4.0\,\mathrm{MHz}$ (\textbf{c}) and $-2.5\,\mathrm{MHz}$ (\textbf{d}). The corresponding paths are illustrated by the insets. All data is binned in intervalls of $0.5\,\mathrm{ms}$. \textbf{e--g,} Absorption images of the atomic momentum distribution, recorded along the $z$--axis after 25\,ms of ballistic expansion with the gas prepared inside the normal phase (\textbf{e}) and the self-organized phases to cavity 1 (\textbf{f}) and 2 (\textbf{g}). Black areas show high atomic densities. The scale bar denotes the length corresponding to a single photon recoil momentum $\hbar k$.}
    \label{fig2}
\end{figure*}

Continuous symmetries can also emerge through symmetry enhancement with competing order parameters---a concept that recently attracted theoretical interest in the context of high-energy physics and cosmology \cite{Lemoine2008, Eichhorn2013}. We create an equivalent situation in a BEC dispersively coupled with two optical cavities and illuminated by a one-dimensional transverse pump lattice (Fig.~\ref{fig1}a). Increasing the coupling to each cavity induces a phase transition to self-organized states, at which atomic ordering is accompanied by intracavity light fields with parity symmetry \cite{Ritsch2013}. These two parity symmetries can be combined to form one U(1) symmetry, similar to constructing the XY model from two Ising models with equal couplings, where the scalar magnetizations form a vector order parameter that exhibits rotational symmetry. Since the position of the atomic ordering along the $x$--axis determines the combination of the cavity field amplitudes (Fig.~\ref{fig1}b), the crystal position and the cavity fields are directly connected (see Methods for the full interference potential). The crystallization is defect-free and homogeneous, because all atoms couple equally to the cavity modes, equivalent to an effective atomic interaction of global range \cite{Asboth2004}.

A microscopic picture of the coupled system is obtained by considering Raman processes between transverse pump and cavity modes which coherently transfer atoms between the motional ground state and excited momentum states (Fig.~\ref{fig1}c) \cite{Ritsch2013, Safaei2015}. Their energies split into $\hbar \omega_+ = 3\hbar \omega_{\mathrm{rec}}$ and $\hbar \omega_- = \hbar \omega_{\mathrm{rec}}$ due to the angle of $60°$ between transverse pump and cavities, with the single photon recoil frequency $\omega_{\mathrm{rec}}$. This description results in the following effective Hamiltonian (see Methods):
\begin{align*}
	\begin{split}
		\hat{\mathcal{H}} /\hbar= \sum_{i \in \{1,2\}}&{\Bigl[-\Delta_i \hat{a}_i^\dagger \hat{a}_i + \omega_+ \cdag{i}{+} \cop{i}{+}  + \omega_- \cdag{i}{-} \cop{i}{-} } \\
			&+ \frac{\lambda}{\sqrt{N}} \left(\hat{a}_i^\dagger  + \hat{a}_i \right) \left( \cdag{i}{+}\cop{0}{} + \cdag{i}{-} \cop{0}{} + h.c. \right)\Bigr] \mathrm{,}
	\end{split}
\end{align*}
where $\hbar$ is the reduced Planck constant and $N$ is the atom number. The atoms are described by creation (annihilation) operators $\cop{0}{}^\dagger$ ($\cop{0}{}$) for the motional ground state as well as for the high- and the low energy states with $\cdag{i}{+}$ ($\cop{i}{+}$) and $\cdag{i}{-}$ ($\cop{i}{-}$), respectively, associated with cavity $i\in \{1,2\}$. The photon fields are denoted by $\hat{a}_i^\dagger$ ($\hat{a}_i$) with detuning $\Delta_i=\omega_\mathrm{p} - \omega_i<0$ between resonance frequency $\omega_i$ and transverse pump laser frequency $\omega_\textrm{p}$.

The interplay of the atoms with each cavity is governed by competition between interaction and kinetic energy scales. While the Raman coupling $\lambda$ favors atomic ordering and the presence of an intracavity field, this comes at the cost of kinetic energy $\omega_{+,-}$ and photonic energy $\Delta_i$. Experimentally, we can access $\lambda$ through the transverse pump lattice depth as well as each $\Delta_i$ by adjusting $\omega_i$. The driven-dissipative character of the experiment only minimally influences the position of the phase boundaries and affects criticalities \cite{Nagy2011, Brennecke2013}. As the decay rates and vacuum Rabi frequencies of the cavities are similar but not equal, the phase boundaries are expected to be displaced by a different amount (see Methods). 

We first explore the phase diagram of the system in order to characterize the competition between the energy terms (see Fig.~\ref{fig2}a, b). Starting point of the experiment is an optically trapped BEC of $^{87}$Rb atoms exposed to an attractive transverse pump lattice potential with wavelength $\lambda_{\mathrm{p}}=785.3\,\mathrm{nm}$ and recoil frequency $\omega_\mathrm{rec}=2\pi\times 3.7\,\mathrm{kHz}$, restricting the motion of the atoms to the $x$--$z$ plane (for experimental details, see Methods). While leaving $\lambda$ constant and independently changing the detunings $\Delta_1$ and $\Delta_2$, we simultaneously record the photons leaking from both cavities, giving real-time access to the intracavity light fields. Due to the concurrence of photonic and atomic excitations, the intracavity light fields allow us to access the degree of atomic ordering.

For each cavity, we observe a buildup of the cavity field at a critical point (Fig.~\ref{fig2}c) indicating the transition to a self-organized state. Three regions are immediately visible: one normal phase without an intracavity field, and two self-organized phases to cavity 1 (SO1) and 2 (SO2). We compare the data with the expected phase boundaries from a mean-field approximation of the Hamiltonian including cavity decay (see Methods) and find good agreement. The phase boundary between the self-organized phases is shifted slightly off-diagonal due to the mismatch in the cavity decay rates. We therefore achieve symmetric coupling to both cavities for cavity 2 slightly closer detuned to the transverse pump frequency than cavity 1.

In order to probe the superfluidity of the atomic cloud, we measure its phase coherence. We suddenly turn off all trapping potentials and allow the atomic wavefunction to expand freely. Subsequently, we perform absorption imaging perpendicular to the cavity plane, see Fig.~\ref{fig2} (e-g). The presence of narrow interference maxima reflecting the initial momentum distribution shows the superfluidity of the cloud \cite{Bloch2008}. In the normal phase, we solely observe the BEC and the momentum peaks at $\pm2\hbar k$ along the direction of the transverse pump with wavenumber $k=2\pi/\lambda_\mathrm{p}$. For finite intracavity field, additional interference maxima appear at momenta of the involved scattering processes depicted in Fig.~\ref{fig1}c. From the interference maxima we conclude that the observed self-organized phases are lattice supersolids with periodicity $d = \lambda_\mathrm{p}/\sin(60°)$. Their broken parity symmetry corresponds to the Hamiltonian being invariant under the unitary transformation $a_i  \rightarrow -a_i$ and accordingly for $\cop{i}{+}$ and $\cop{i}{-}$ for each cavity $i\in \{1,2\}$. The lattice supersolid for cavity 1 (2) corresponds to atomic density modulation along the $x$--axis forming at discrete positions 0 or $d/2$ ($d/4$ or $3d/4$).

\begin{figure}[t]
    \includegraphics[width=\columnwidth]{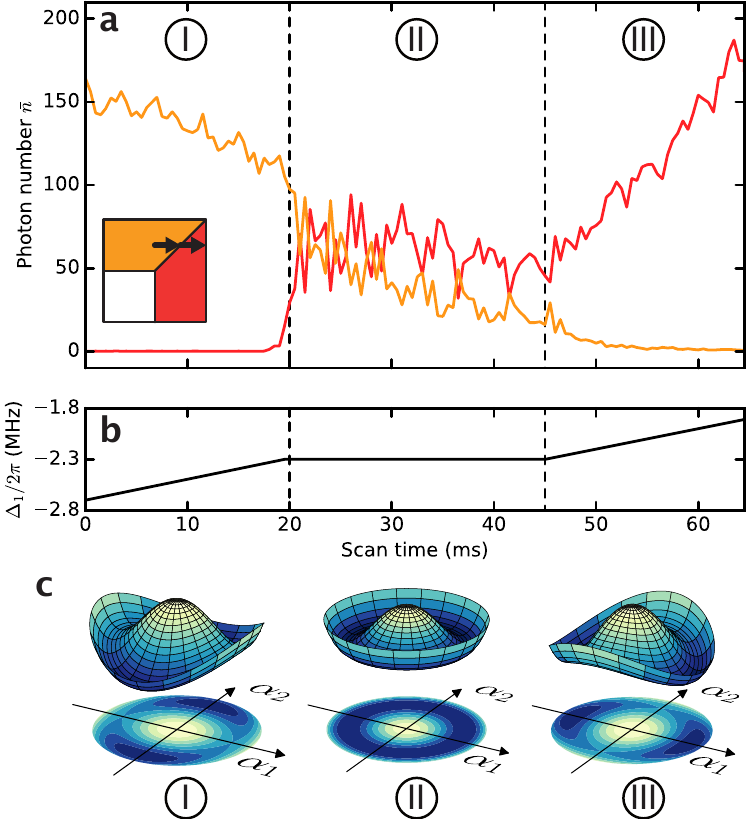}
    \caption{{\bf Emergence of a doubly self-organized phase.} \textbf{a,} Mean intracavity photon numbers (binned in intervalls of $0.5\,\mathrm{ms}$) for the frequency ramp shown in \textbf{(b)} for cavity 1 (red) and 2 (yellow) at constant transverse pump lattice depth $38(1)\,\hbar\omega_\mathrm{rec}$. The simultaneous presence of photons in both cavities signals the transition to a doubly self-organized phase. \textbf{b,} Detuning ramp through the phase diagram. At constant $\Delta_2/2\pi=-2.2\,\mathrm{MHz}$, $\Delta_1$ is ramped from far-detuned to reaching the phase boundary, held there for $25\,\mathrm{ms}$ and subsequently ramped closer to resonance. \textbf{c,} The mean-field energy as a function of cavity field amplitudes $\alpha_1$ and $\alpha_2$ is qualitatively displayed for three different regions in the phase diagram: inside the self-organized phase to cavity 2 (I) and 1 (III), and on the phase boundary in between (II).}
    \label{fig3}
\end{figure}

The crucial feature of the phase diagram is the discontinuity of the order parameters between the SO1 and the SO2 phases (Fig.~\ref{fig2}d). We investigate this situation more closely by probing the system on the phase boundary (Fig.~\ref{fig3}a, b). After preparing the atoms in the SO2 phase, we change the detuning $\Delta_1$ until the phase boundary is reached. There we keep the detunings fixed and monitor the evolution of both cavity light fields. We observe finite mean intracavity photon numbers in both cavities, signaling a new type of self-organization. This additional phase is visible only in a narrow region of $\delta\Delta_1\sim100\,\mathrm{kHz}$ in the phase diagram. During the evolution, anticorrelations in the cavity light fields are visible on the ms time scale, which we quantify with a Pearson correlation coefficient of $-0.82(9)$ extracted from fourteen realizations of the same experiment.

We interpret the emergence of the doubly self-organized phase in terms of the mean-field energy as a function of the cavity field amplitudes $\alpha_1$ and $\alpha_2$, as illustrated in Fig.~\ref{fig3}c. Starting in the SO2 phase, the mean-field energy exhibits two minima on the $\alpha_2$--axis corresponding to the underlying parity symmetry. These minima are rotated to the $\alpha_1$--axis when entering the SO1 phase (region III). Only at the point in between, the ground state manifold extends to a circle that connects both axes (region II), thereby realizing a U(1) symmetry. This is also visible from the Hamiltonian, which is invariant under the unitary transformation $(\hat{a}_1, \hat{a}_2) \rightarrow \left(\hat{a}_1\cos \theta + \hat{a}_2\sin \theta, -\hat{a}_1\sin \theta + \hat{a}_2\cos \theta\right)$, and accordingly for $(\cop{1}{+},\cop{2}{+})$ and $(\cop{1}{-},\cop{2}{-})$, for any rotation angle $\theta$ (see Methods). A low-energy evolution on the ground state manifold results in the observed anticorrelated signals in the mean intracavity photon numbers $\overline{n}_i=|\alpha_i |^2$, corresponding to a spatial displacement of the density modulation along the $x$--axis. 

\begin{figure}[t]
    \includegraphics[width=\columnwidth]{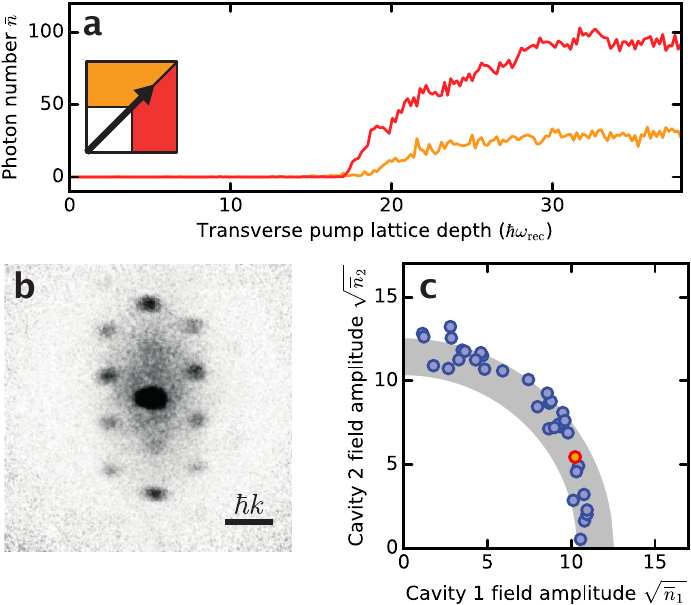}
    \caption{{\bf Breaking continuous translational symmetry.}  \textbf{a,} Examplary traces of the mean photon numbers in cavity 1 (red) and 2 (yellow) as a function of transverse pump lattice depth for a ramp time of $100\,\mathrm{ms}$ at constant $\Delta_1 = -2.1\,\mathrm{MHz}$ and $\Delta_2 = -2.0\,\mathrm{MHz}$. The ramp corresponds to a single diagonal scan in the phase diagram (see inset). Traces are binned in intervalls of $0.5\,\mathrm{ms}$. \textbf{b,} Absorption image of the atomic momentum distribution, recorded along the $z$--axis after $25\,\mathrm{ms}$ ballistic expansion. Black areas show high atomic densities. The eight atomic momentum modes associated to the scattering of photons from the pump to the cavities are visible. \textbf{c,} Field amplitudes for 35 different experimental realizations of (\textbf{a}), deduced from the intracavity photon numbers averaged over $5\,\mathrm{ms}$ at the end of the ramp. The red data point denotes the ramp shown in (\textbf{a}). The data reveal the continuous translational symmetry which is broken in the supersolid. The grey shaded area contains the systematic uncertainty from the calibration of the intracavity photon numbers.}
    \label{fig4}
\end{figure}

We use the connection between crystal position and intracavity fields to further characterize the ground state symmetry of the doubly self-organized phase. To this end, we ramp up the transverse pump power with symmetric couplings to the two cavities, resulting in a more adiabatic preparation of the system. When entering the doubly self-organized phase, we observe finite intracavity photon numbers in both cavities (Fig.~\ref{fig4}a) and interference maxima at momenta associated with scattering processes for both cavities are visible in absorption imaging (Fig.~\ref{fig4}b). Repeating this experiment, we observe that the cavity field amplitudes distribute on a quarter circle (Fig.~\ref{fig4}c), revealing a high ground state degeneracy of the system. Whilst the U(1) symmetry includes the sign of the cavity field amplitudes, our photon detection only allows to measure their magnitude. Therefore the full circular symmetry is folded to the positive quadrant. As the combination of the intracavity field amplitudes determines the position of the atomic density modulation along the $x$--axis, we conclude that in each realization the superfluid crystallizes to a different position. This realizes a supersolid phase, as it exhibits both superfluid phase-coherence and broken continuous translational symmetry. 

An interesting extension to our work would be to project a speckle potential on the atomic cloud, which lifts the ground-state degeneracy in a controlled way. This permits to realize supersolidity in the presence of disorder and to study dynamics in a quasi-degenerate ground-state manifold \cite{Gopalakrishnan2009, Strack2011}. Beyond supersolidity, our setup can also be interpreted as identical two-level systems coupled with two quantized light fields, therefore providing access to a new class of quantum optical models breaking a continuous symmetry \cite{Fan2014, Baksic2014}. 

We thank E. Demler, J.~Larson and B.~P. Venkatesh for insightful discussions on U(1) symmetries in optical cavities and M.~Lee for contributions in the building stage of the experimental apparatus. We acknowledge funding from Synthetic Quantum Many-Body Systems (European Research Council advanced grant) and the EU Collaborative Project TherMiQ (Grant Agreement 618074), and also SBFI support for Horizon2020 project QUIC and SNF support for NCCR QSIT and DACH project `Quantum Crystals of Matter and Light'.


\clearpage

\section*{Methods}

\setcounter{equation}{0}
\setcounter{figure}{0}
\renewcommand{\figurename}{Extended Data Figure}
\renewcommand{\tablename}{Extended Data Table}

\subsection*{Preparation of the Bose-Einstein condensate}
After optically transporting a cold thermal cloud of $^{87}$Rb atoms along the $x$--axis [31] into the cavity setup, we optically evaporate to an almost pure Bose-Einstein condensate with $N=1.50(3)\times 10^5$ atoms in a dipole trap, which is formed by two orthogonal laser beams at a wavelength of $1064\,\mathrm{nm}$ along the $x$-- and $y$--axes. The final trapping frequencies are $(\omega_x, \omega_y, \omega_z)=2\pi \times (88(1), 76(3), 154(1))\,\mathrm{Hz}$, resulting in Thomas-Fermi radii of $(R_x,R_y,R_z)=(7.9(1), 9.2(2), 4.5(1))\,\mu\mathrm{m}$. We subsequently expose the atoms to an attractive one-dimensional lattice potential along the $y$--direction by linearly ramping up the transverse pump beam to a lattice depth of $38(1)\,\hbar \omega_{\mathrm{rec}}$ within $50\,\mathrm{ms}$. The mirror retroreflecting the transverse pump beam is positioned in vacuum at a distance of $8.6\,\mathrm{mm}$ from the atomic position. The beam has a $1/e^2$ radius of $(w_x,w_z)=(35(3),45(3))\mu\mathrm{m}$ and is polarized along $\hat{z}$, parallel to gravity. The wavelength is set to $\lambda_\mathrm{p}=785.3\,\mathrm{nm}$, far red-detuned with respect to the atomic $D_2$ line.  

\subsection*{Two crossed optical cavities} We mounted the mirrors forming the Fabry-Perot cavities as close as possible to each other to achieve large vacuum Rabi rates. For two crossing cavities, this required us to specifically machine the substrates before glueing them in place. The two cavities have comparable single-atom vacuum Rabi frequencies of $(g_1, g_2) = 2\pi\times (1.95(1), 1.77(1))\,\mathrm{MHz}$, and decay rates $(\kappa_1, \kappa_2) = 2\pi\times (147(4), 800(11))\,\mathrm{kHz}$. The cavity modes intersect at an angle of 60$^\circ$ and have $1/e^2$ radii of $(49(1),50(1))\,\mu\mathrm{m}$. We position the atoms vertically in between the two mode axes, therefore at a distance of $8(2)\,\mu\mathrm{m}$ from each mode center. For each cavity, we individually set the frequency of a longitudinal mode closely detuned with respect to the transverse pump frequency by stabilizing the cavity lengths with a weak stabilization laser beam at $830.4\,\mathrm{nm}$. The resulting additional intracavity lattice potential is incomensurate with the transverse pump wavelength and has a depth of $0.1(1)\,\hbar\omega_{\mathrm{rec}}$, negligible compared to the self-organisation lattice depth in the organized phases of typically $2$--$4\,\hbar\omega_\mathrm{rec}$. Long-term frequency stability between the transverse pump laser and the stabilization laser of around $50\,\mathrm{kHz}$ is achieved by locking all of them simultaneously to a passively stable transfer cavity. Single-photon counting modules are used to detect photons leaving the cavities. 

\subsection*{Lattice calibrations} The lattice depths of the transverse pump and the intracavity lattices are calibrated performing Raman-Nath diffraction on the cloud. The calibrated intracavity photon number can then be deduced from the vacuum Rabi frequencies of the cavities. We extract overall intracavity photon detection efficiencies of $(\eta_1, \eta_2)=(9.7(4)\%,2.0(1)\%)$ with relative systematic uncertainties of $8\%$.

\subsection*{Cavity-induced spin transitions} The BEC is prepared in the $\ket{F, m_F}=\ket{1,-1}$ state with respect to the quantization axis along $\hat{z}$.  The birefringences for the two cavities between the $H$ and $V$ eigenmodes are $(4.5, 4.8)\,\mathrm{MHz}$. Whenever the resonance condition for a two-photon process involving pump and cavities is met, we observe collective Raman transitions between different Zeeman sublevels accompanied by macroscopic occupation of the cavity modes. In order to suppress such spin changing scattering processes, all data was taken at large offset field of $B_z=34\,\mathrm{G}$, creating a Zeeman level splitting large compared to $\Delta_i$ and the birefringences. 

\subsection*{Effective Hamiltonian} We start with the following many-body Hamiltonian in a frame rotating at the pump laser frequency \cite{Ritsch2013}:
\begin{equation}
\begin{aligned}
\mathcal{H_{\mathrm{mb}}} = & \sum_{i=1,2}\Bigl(-\hbar \Delta_i \hat{a}_i^\dagger \hat{a}_i \Bigl)+ \iint\limits_A \hat{\Psi}^\dagger(x,y) \biggl[ \frac{p_x^2+p_y^2}{2m}\\
& + \sum_{i=1,2}\Bigl(\hbar \eta_i (\hat{a}_i^\dagger+\hat{a}_i) \cos \left({\bf k}_\mathrm{p}\cdot {\bf r}+ \phi \right) \cos\left({\bf k}_i \cdot {\bf r}\right) \\
& + \hbar U_i \cos^2 \left( {\bf k}_i\cdot {\bf r}\right)  \hat{a}_i^\dagger \hat{a}_i\Bigl)\\
& + \hbar U_\mathrm{p} \cos^2 \left( {\bf k}_\mathrm{p}\cdot {\bf r} + \phi \right) \biggr]\hat{\Psi}(x,y) \,\mathrm{d}x\,\mathrm{d}y, 
\label{eq:theory_origH}
\end{aligned}
\end{equation}
where the index $i\in\{1,2\}$ labels the two cavities, $\hbar$ is the reduced Planck constant, $\phi$ is the spatial phase of the transverse pump lattice, $\eta_i=\frac{\Omega_\mathrm{p}g_i}{\Delta_\mathrm{a}}$, $U_i=\frac{g_i^2}{\Delta_\mathrm{a}}$ and $U_p$ are the potential depths of the cavity and the transverse pump fields, respectively, and the integration runs over the area $A$ of the Wigner-Seitz cell. The standing-wave transverse pump beam at frequency $\omega_\mathrm{p}$ with Rabi frequency $\Omega_\mathrm{p}$ is red-detuned by $\Delta_\mathrm{a}=\omega_\mathrm{p}-\omega_\mathrm{a}$ from the atomic resonance at frequency $\omega_\mathrm{a}$. It is oriented along the $y$--axis with a wave-vector ${\bf k}_\mathrm{p}=k\hat{y}$. The cavities $i\in\{1,2\}$ at frequencies $\omega_i $ are detuned by $\Delta_i=\omega_\mathrm{p} - \omega_i$ from the transverse pump frequency and are described by modes $\hat{a}_i$ and wavevectors ${\bf k}_i= k\sin (60^\circ)\hat{x} + (-1)^i \cos(60^\circ)\hat{y}$. $\hat{\Psi}\dagger(x,y)$ ($\hat{\Psi}(x,y)$) are the atomic field operators creating (annihilating) a particle at position $(x,y)$. We neglect any cavity decay rate in this part of our formalism as well as the cavity-cavity interference term which is negligible for the choice of our experimental parameters. 

The first line of the Hamiltonian describes the energy of the photon fields in the cavities and the kinetic energy of the atoms. The remaining terms take into account the cavity-pump interference terms for the two cavities, the cavity lattices, and the pump lattice. The transverse pump lattice beam at wavelength $\lambda_\mathrm{p}=\frac{2\pi}{k}$ is far red-detuned with respect to atomic resonance, but closely detuned by $\Delta_i$ to the two cavity resonances. Therefore, photons from the transverse pump can be scattered into a cavity mode and back via off-resonant Raman processes. These two-photon processes coherently couple the zero momentum state of the BEC $\ket{\hbar k_x, \hbar k_y}=\ket{0,0}$ to the eight momentum states $\ket{\hbar k_x, \hbar k_y}=\ket{\pm\mathrm{\bf \hbar k_p \pm \hbar k}_i}$ which are sketched in Fig.~1c. We neglect scattering processes between the two cavities, and between excited atomic momentum states, as their amplitudes are negligible for our experimental parameters. Since the cavities and the transverse pump are not orthogonal, these eight momentum states group in high-energy and low-energy states with energies $\hbar \omega_+ = 2\hbar \omega_{\mathrm{rec}} (1+\cos(60°)) = 3\hbar \omega_{\mathrm{rec}}$ and $\hbar \omega_- = 2\hbar \omega_{\mathrm{rec}} (1-\cos(60°)) = \hbar \omega_{\mathrm{rec}}$. As ansatz for the atomic field operator we choose 
\begin{equation}
\hat{\Psi} = \Psi_{0}{} \hat{c}_0+ \sum_{i=1,2}\bigl(\Psi_{i-} \hat{c}_{i-}+ \Psi_{i+} \hat{c}_{i+}\bigl),
\label{eq:psi}
\end{equation}
where $\Psi_{0}{} =\sqrt{\frac{2}{A}}$ represents the BEC zero-momentum mode, and the functions $\Psi_{i \pm} = \sqrt{ \frac 2 A } \cos\left[ \left( {\bf k}_\mathrm{p} \pm {\bf k}_i \right) \cdot {\bf r} \right]$ represent the atomic modes with momentum imprinted by one of the scattering processes at high or low energy into one of the two cavities.

After carrying out the integration in Eq.~\ref{eq:theory_origH} using Eq.~\ref{eq:psi}, we obtain the effective Hamiltonian
\begin{equation*}
\begin{aligned}
\hat{\mathcal{H}} =\sum_{i=1,2}\Bigl[&-\hbar\Delta_i\hat{a}^{\dagger}_i\hat{a}_i+\hbar\omega_{+}\hat{c}^{\dagger}_{i+}\hat{c}_{i+}+\hbar\omega_{-}\hat{c}^{\dagger}_{i-}\hat{c}_{i-}\\
&+\frac{\hbar\lambda_i}{\sqrt{N}}\Bigl(\hat{a}^{\dagger}_i+\hat{a}_i\Bigl)\Bigl(\hat{c}^{\dagger}_{i+}\hat{c}_{0}+\hat{c}^{\dagger}_{i-}\hat{c}_{0}+h.c.\Bigl)\Bigl],
\end{aligned}
\end{equation*}
where we have introduced the Raman coupling $\lambda_i=\frac{\eta_i\sqrt{N}}{2\sqrt{2}}$ that can be controlled via $\eta_i=-\frac{\Omega_\mathrm{p}g_i}{\Delta_a}$. The dispersive shift $N U_0/2$ is very similar for both cavities, and therefore can be absorbed into $\Delta_i$. Other optomechanical terms can be discarded for our parameters. Neglecting a difference between the two Raman couplings, as in the Hamiltonian of the main text, only leads to a small correction because the vacuum Rabi frequencies $g_i$ are similar.

\subsection*{Critical couplings}
Self-organization to a single cavity occurs when the coupling $\lambda_i$ crosses the critical coupling $\lambda^{\mathrm{cr}}_i = \sqrt{-\Delta_i\overline{\omega}/4}$, with $\overline{\omega}^{-1}=\omega_+^{-1}+\omega_-^{-1}$. This crossing is obtained by either changing $\lambda_i$ or $\Delta_i$. The phase boundary between the SO1 and the SO2 phase is identified by the condition $\frac{\lambda_1}{\lambda^{\mathrm{cr}}_{1}}=\frac{\lambda_2}{\lambda^{\mathrm{cr}}_{2}}$, where the coupling to both cavities is symmetric. Including dissipation rates $\kappa_i$ for the cavity fields changes the critical couplings into $\lambda^{\mathrm{cr}}_i = \sqrt{-(\Delta_i^2+\kappa_i^2)/\Delta_i\overline{\omega}}$.

\subsection*{Symmetries of the Hamiltonian}
The Hamiltonian can be read as the sum of two formally identical Hamiltonians for the two cavities, $\hat{\mathcal{H}} =\sum_{i=1,2}\hat{\mathcal{H}} _i$. 

Both $\hat{\mathcal{H}}_i$ individually manifest parity symmetry, generated by the operator $\hat{C_i}=\hat{a_i}^\dagger\hat{a_i} + \sum_{s=\pm}\hat{c}^{\dagger}_{is}\hat{c}_{is}$. They stay unchanged upon the simultaneous transformation $(\hat{a_i},\hat{c}_{i\pm})\rightarrow-(\hat{a_i},\hat{c}_{i\pm})$  on the photonic and atomic field operators [32]. This symmetry is broken at the phase transition. The choice of sign of the field operators now corresponds to a choice of 0 or $\pi$ for the phase of the light field in cavity $i$, which is equivalent to atoms crystallizing on odd or even sites of a chequerboard lattice with rhomboid geometry.

For $\lambda_1=\lambda_2$ and $\Delta_1=\Delta_2$, the Hamiltonian $\hat{\mathcal{H}}$ exhibits a $U(1)$ symmetry instead: it is possible to perform a simultaneous rotation by an arbitrary angle $\theta$ in the space of the cavity field and atomic field operators that leaves the Hamiltonian $\hat{\mathcal{H}}$ unchanged. The transformation acts in the following way on the operators:
\begin{equation*}
\begin{aligned}
\hat{a}_1 & \rightarrow \hat{a}_1 \cos \theta - \hat{a}_2 \sin\theta \\
\hat{a}_2 & \rightarrow \hat{a}_1 \sin \theta + \hat{a}_2 \cos\theta \\
\hat{c}_{1\pm} & \rightarrow \hat{c}_{1\pm} \cos \theta - \hat{c}_{2\pm} \sin\theta \\
\hat{c}_{2\pm} & \rightarrow \hat{c}_{1\pm} \sin \theta + \hat{c}_{2\pm} \cos\theta.
\end{aligned}
\end{equation*}
It shifts photons between the two cavities while simultaneously redistributing the momentum excitations accordingly. This produces the circle on the $\alpha_1$-$\alpha_2$-plot in Fig.~4c. The corresponding generator $\hat{C}$ of the symmetry $\hat{U}(\theta)=e^{i\theta\hat{C}}$ is the Hermitian operator 
\begin{equation}
\hat{C}=-i\Bigl[\hat{a}_1^\dagger \hat{a}_2 - \hat{a}_2^\dagger \hat{a}_1 + \sum_{s=\pm}\Bigl(\hat{c}^{\dagger}_{1s} \hat{c}_{2s} - \hat{c}^{\dagger}_{2s} \hat{c}_{1s}\Bigl)\Bigl]. 
\end{equation}
It satisfies $[\hat{C},\hat{\mathcal{H}}]=0$, and, as a direct consequence, the Hamiltonian $\hat{\mathcal{H}}$ stays unchanged for any $\theta$ under the symmetry $\hat{U}$, i.\,e. $\hat{U} \hat{\mathcal H} \hat{U}^\dagger = \hat{\mathcal H}$. This symmetry is spontaneously broken at the phase transition. 

\begin{figure}[t]
 \includegraphics[width=\columnwidth]{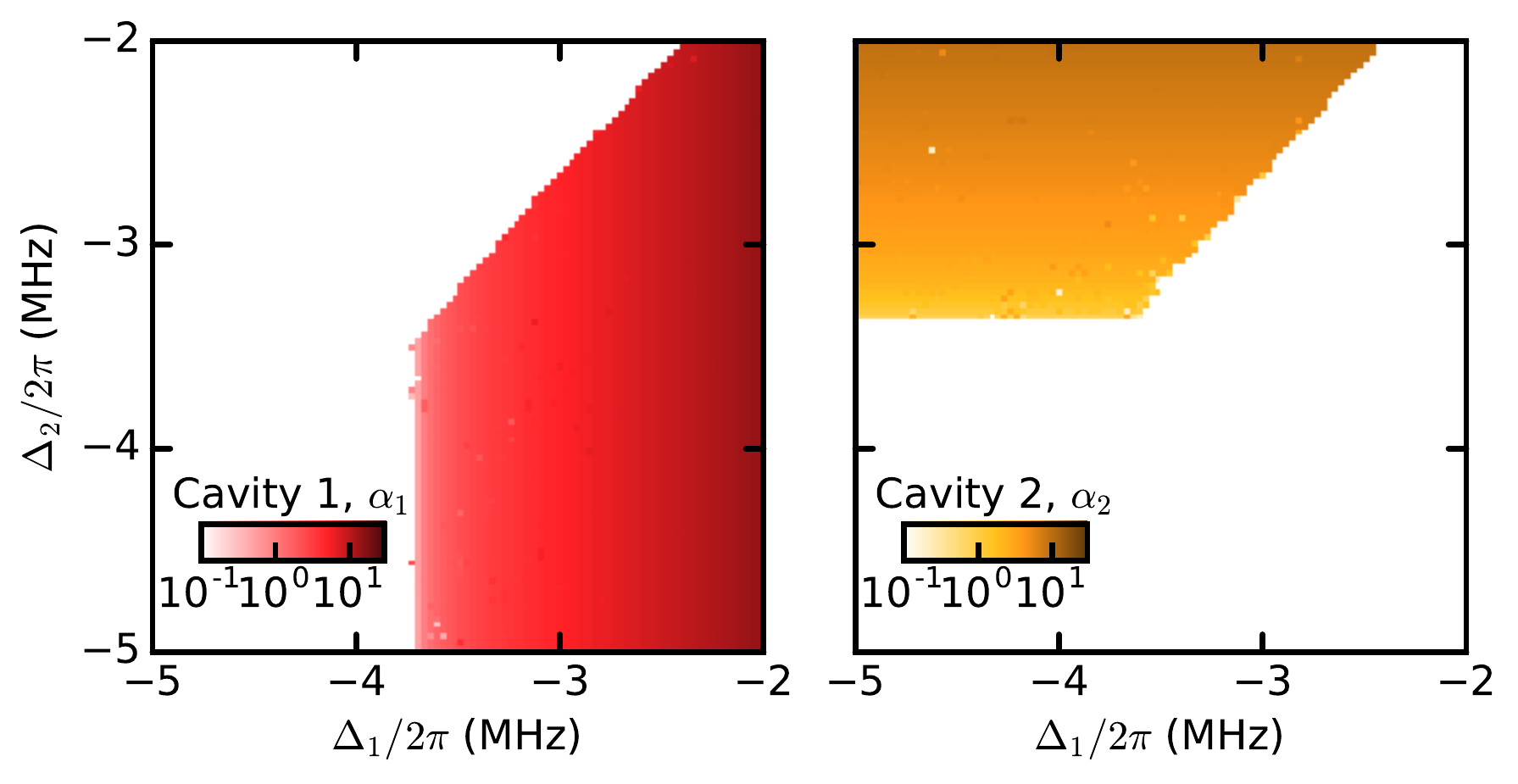}
\caption{\textbf{Mean-field solution of the phase diagram} Order parameters $\alpha_1$ and $\alpha_2$ as a function of the pump--cavity detunings $\Delta_1$ and $\Delta_2$. The Raman coupling $\lambda$ is fixed by the experimental value of the transverse pump lattice depth of $38\hbar\omega_\mathrm{rec}$.  The mean-field model includes cavity decay, different vacuum Rabi frequencies $g_i$ and the transverse pump potential.}
\label{fig:mf_pd}
\end{figure}

\subsection*{Mean-field solution} We can obtain the ground state of the effective Hamiltonian by performing a mean-field expansion around the expectation value of each operator and numerically solving the resulting mean-field Hamiltonian [33]. 
To get a direct comparison with the experimentally measured data, we plot the order parameters $\alpha_1$ and $\alpha_2$ as a function of the two detunings $\Delta_i$ (see Extended Data Fig.~\ref{fig:mf_pd}. The calculation includes cavity decay, the influence of the transverse pump potential and atom-atom contact interactions.

\subsection*{Supersolid potential}
The self-consistent potential in the supersolid phase is formed from the interference between the transverse pump field and the two cavity fields. We derive here the relation between the spatial position of the optical lattice structure and the ratio between the coherent fields $\alpha_1$ and $\alpha_2$ in each cavity. 

\begin{figure}[b]
 \includegraphics[width=\columnwidth]{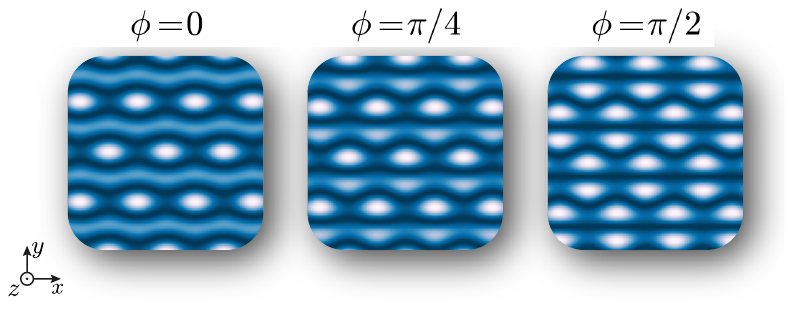}
\caption{\textbf{Lattice geometries for different choices of the the phase $\phi$ of the transverse pump field for balanced cavity fields.} The atoms are cut into 1D lines by the strong transverse pump field. On top, spatial distributions between triangular ($\phi=0$) and hexagonal ($\phi=\pi/2$) can form through the interference between the cavity light fields and the transverse pump, depending on the phase $\phi$ of the latter.}
\label{fig:potential_phi}
\end{figure}

The full potential landscape for the atoms is given by the coherent superposition of the transverse pump field and the two cavity fields,
\begin{equation}
U({\bf r}) = \left( \sum\limits_{i\in\{\mathrm{p},1,2\}} \Omega_i \cos ({\bf k}_i \cdot {\bf r} + \phi_i) \right)^2,
\end{equation}
where $\Omega_i^2=U_i$ are the potential depths created by each field. We set $\phi_1=\phi_2=0$ by choosing the origin of the coordinate system appropriately. The atomic spatial distribution is then determined by the phase $\phi_p\equiv\phi$ of the transverse pump standing wave, which we can change via a piezo--electric actuator attached to the retroreflecting transverse pump mirror. For our experimental parameters $U_\mathrm{p}\gg U_1, U_2$ such that the atoms are separated into two-dimensional layers in the $x$--$z$ plane at $ky +\phi = \pi n, n\in\mathbb Z$, where $k=2\pi/\lambda_\mathrm{p}$. We can tune the spatial phase $\phi$ of the transverse pump standing wave by a piezoelectric actuator at the retroreflecting mirror, resulting in triangular ($\phi = 0$) and hexagonal ($\phi = \pi/2$) lattice geometries, c.f.~Extended Data Fig.~\ref{fig:potential_phi}.

\begin{figure}[t]
 \includegraphics[width=\columnwidth]{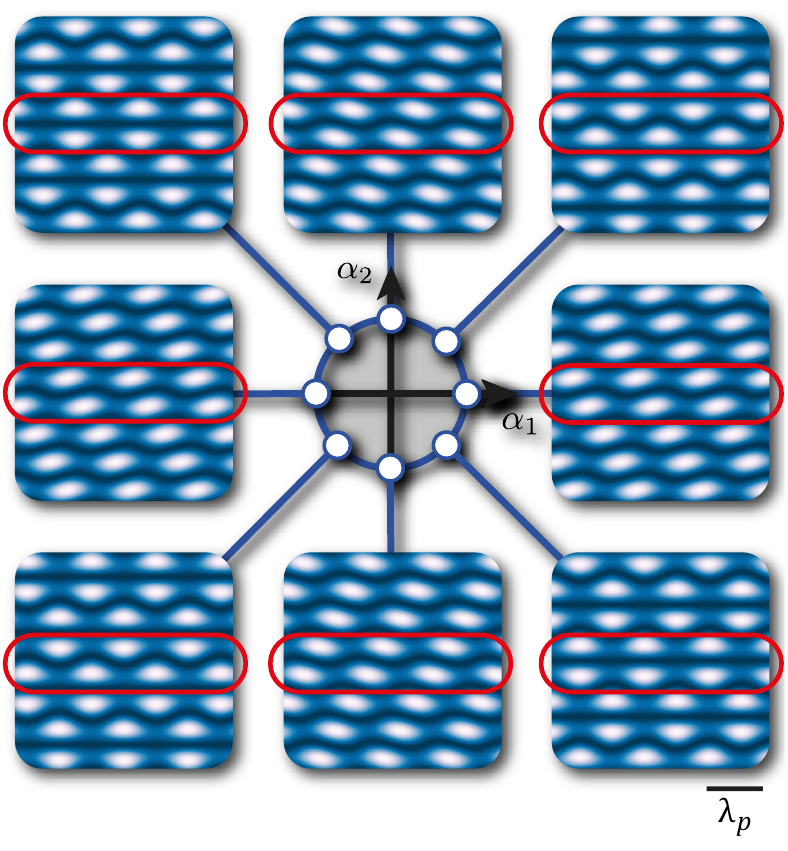}
\caption{\textbf{Dependence of the lattice structure on the cavity field amplitudes.} The ground-state manifold for equal couplings and detunings in $\mathcal{H}$ is a circle in the space of the cavity fields $\alpha_1$ and $\alpha_2$. For each combination of fields, the interference potential in Eq.~\ref{eq:Ur} between transverse pump and cavity fields for $\phi=\pi/2$ will have its minima at different positions. Following the circle clockwise, every second line moves left (top highlighted line) while the others move right (bottom highlighted line).}
\label{fig:potentials}
\end{figure}

Within one layer we get
\begin{align}
	\begin{split}
		U(x) = & (\Omega_\mathrm{p} \cos(2\phi) \\
		& + \Omega_\mathrm{c} [(\cos\theta+\sin\theta) \cos(\phi/2)\cos(\sqrt3\pi x) \\
		& + (\cos\theta-\sin\theta)\sin(\phi/2)\sin(\sqrt3\pi x) ])^2,
	\end{split}
\end{align}
where $\Omega_1=\Omega_\mathrm{c}\cos\theta$ and $\Omega_2=\Omega_c\sin\theta$ with $\theta$ corresponding to the position on the $\alpha_1$--$\alpha_2$--circle as before. This describes a lattice whose position depends on $\theta$, unless $\phi=0$, where only the lattice depth is modified. The lattice depth has a $\theta$--modulation that disappears as $\phi$ approaches $\pi/2$, in which case Eq.~\ref{eq:Ur} simplifies to $[-\Omega_\mathrm{p}+\Omega_\mathrm{c}\cos(\theta+\sqrt3\pi x-\pi/4)]^2$. We choose $\phi\approx\pi/2$ in our experiments such that in the broken $U(1)$ symmetry each realization of cavity fields corresponds to a different translation, as shown in Extended Data Fig. \ref{fig:potentials}. Neighbouring layers move in opposite directions, so that the translation is staggered. 

\end{document}